\documentclass[aps,prl,reprint,twocolumn,nofootinbib,amsmath,amssymb,floatfix,preprintnumbers,nobibnotes,longbibliography]{revtex4-2}

\usepackage{amsmath,amssymb,mathtools,bm,graphicx}
\usepackage{xcolor}
\usepackage{xparse}

\newcommand{\tr}{\operatorname{tr}}
\newcommand{\diag}{\operatorname{diag}}
\newcommand{\dmod}{\mathrm{mod}\,}

\renewcommand{\Im}{\mathrm{Im}}
\DeclarePairedDelimiter{\abs}{\lvert}{\rvert}
\NewDocumentCommand{\order}{s m}{%
  \mathcal O\IfBooleanTF{#1}{\bigl(#2\bigr)}{\bigl(#2\bigr)}%
}

\graphicspath{{fig/}}
\allowdisplaybreaks[3]
\begin{document}

\preprint{RESCEU-22/26}

\title{Imaginary Rotation Breaks Charge Conjugation in Hot QCD}

\author{Rin Takada}
\email{takada-rin@resceu.s.u-tokyo.ac.jp}
\affiliation{Research Center for the Early Universe (RESCEU), Graduate School
of Science,
The University of Tokyo, 7-3-1 Hongo, Bunkyo, Tokyo 113-0033, Japan}
\date{\today}

\begin{abstract}
Starting from the identity
$\mathrm e^{2\pi\mathrm i\hat J_z}=(-1)^{\hat F}=\mathrm e^{\mathrm i\pi\hat Q}$
on the color-singlet physical state space, we derive a
density-operator identity that ties imaginary rotation to an imaginary
quark chemical potential,
$\hat\rho(\tilde\Omega_I+2\pi,\theta_q)=\hat\rho(\tilde\Omega_I,\theta_q+\pi)$.
Hence imaginarily rotating $SU(3)$ QCD at
$(\tilde\Omega_I,\theta_q)=(2\pi,0)$ is mapped exactly onto the
Roberge-Weiss (RW) point $(0,\pi)$, where charge conjugation $C$ is
spontaneously broken above the RW endpoint---a conclusion independent
of any model or approximation.
Minimizing the one-loop effective potential on the rotation axis, we
obtain analytically, for three massless flavors, the second-order
transition at $\tilde\Omega_{I,C}/\pi=(22-2\sqrt{22})/9$,
and a subsequent first-order transition at
$\tilde\Omega_{I,{\rm lock}}/\pi=2/3+2\sqrt{111}/27$.
At smaller $\tilde\Omega_I$, a continuous degeneracy of the massless
one-loop approximation leaves the realization of $C$ undecided; a
finite strange-quark mass lifts it.
We also state how the exact relations and the on-axis predictions can
be tested in lattice QCD.
\end{abstract}

\maketitle

\emph{Introduction.}---Rotation does not tell matter from antimatter: the
angular velocity is
invariant under charge conjugation $C$, and at first sight nothing
permits rotation to break this symmetry.
Nevertheless, as we show in this Letter, imaginary rotation alone,
without any imaginary quark chemical potential, maps hot QCD exactly
onto a phase with spontaneously broken $C$.

Rotation has become a standard deformation of hot QCD: the enormous
vorticity of non-central heavy-ion collisions, observed as the global
polarization of hyperons
~\cite{Liang:2004ph,STAR:2017ckg,Becattini:2020ngo},
has prompted extensive studies of rotating QCD matter
~\cite{Chen:2015hfc,Jiang:2016wvv,Chernodub:2016kxh,Ebihara:2016fwa}.
For a real angular velocity the Euclidean action is complex, so
lattice QCD relies on special setups or on analytic continuation from
imaginary angular velocity
~\cite{Yamamoto:2013zwa,Braguta:2020biu,Braguta:2023yjn}.

A theory with imaginary angular velocity is, however, interesting in its own
right: at high temperature, a one-loop calculation reveals a perturbatively
confined phase on the rotation axis~\cite{Chen:2022smf}; with dynamical quarks
a $4\pi$ periodicity arises in the imaginary angular
velocity~\cite{Chen:2023cjt,Chen:2024tkr}, and inhomogeneous structures away
from the axis have been analyzed~\cite{Chen:2024tkr}.

Under an imaginary quark chemical potential $\theta_q$,
the Roberge-Weiss (RW) periodicity holds~\cite{Roberge:1986mm}, and on
the RW transition lines the residual $\mathbb Z_2$ symmetry---charge
conjugation combined with a gauge transformation involving a center
element---is spontaneously broken above the RW endpoint temperature $T_{\rm
RW}$, as established by lattice QCD
~\cite{DElia:2009qz,deForcrand:2010he,Bonati:2010gi,Bonati:2016RW}
and by effective models
~\cite{Sakai:2008py,Kouno:2009rw,Fukushima:2017csk}.

The point $\theta_q=\pi$ renders the quarks periodic in Euclidean
time and, for odd $N_c$, lies on these transition lines.
That such periodic fundamental-representation fermions break $C$
spontaneously was first discussed in small volumes~\cite{vanBaal:1988va}
and is known at one loop~\cite{MyersOgilvie:2009} and on the
lattice in $S^1$-compactified QCD, together with the accompanying
baryonic current~\cite{DeGrandHoffmann:2007,LuciniPatellaPica:2007},
and has recently been restated in the RW context at odd $N_c$
~\cite{DEliaNacciZambello:2026}.
The periodic-fermion phase and the $4\pi$ periodicity are thus both
known.
The novelty of this Letter is the bridge: we identify imaginarily
rotating QCD, exactly and at the level of the density operator, with
this known $C$-breaking phase, and we determine analytically the second-order
transition point of $C$ breaking and the first-order lock-in point
that appear on the rotation axis as the imaginary angular velocity is
varied.

\emph{An identity connecting rotation to quark number.}---We
write $\Omega\eqqcolon\mathrm i\Omega_I$,
$\tilde\Omega_I\coloneqq\beta\Omega_I=\Omega_I/T$, and
$\mu_q\eqqcolon\mathrm i\theta_qT$ for the imaginary angular velocity,
its dimensionless version, and the imaginary quark chemical potential.
Imaginary rotation is free of the causality condition of real rotation,
so no finite-radius boundary is required and the thermodynamic limit
can be taken~\cite{Chen:2022smf}.
Let $\hat H$ be the Hamiltonian, $\hat J_z$ the $z$ component of the
total angular momentum, and $\hat Q$ the net quark number operator; we
assume $[\hat H,\hat J_z]=[\hat H,\hat Q]=0$.
The operator defining the thermal ensemble is
\begin{align}
\hat\rho(\tilde\Omega_I,\theta_q)
=\mathrm e^{-\beta\hat H}
\mathrm e^{\mathrm i\tilde\Omega_I\hat J_z}
\mathrm e^{\mathrm i\theta_q\hat Q},
\label{eq:imaginary-rotation-C:partition-function}
\end{align}
and the partition function is
$Z(\tilde\Omega_I,\theta_q)=\tr\hat\rho(\tilde\Omega_I,\theta_q)$,
the trace being taken over the color-singlet physical state space
satisfying the Gauss law.

A $2\pi$ rotation multiplies bosonic states by $+1$ and fermionic
states by $-1$: $\mathrm e^{2\pi\mathrm i\hat J_z}=(-1)^{\hat F}$.
In QCD the fermions are the quarks and antiquarks, so
$\hat F=\hat N_q+\hat N_{\bar q}\equiv\hat N_q-\hat N_{\bar q}
=\hat Q\pmod 2$; hence, on the physical state space,
\begin{align}
\mathrm e^{2\pi\mathrm i\hat J_z}
=(-1)^{\hat F}
=\mathrm e^{\mathrm i\pi\hat Q}.
\label{eq:imaginary-rotation-C:operator-identity}
\end{align}
That $-1\in U(1)_{\rm B}$ acts as the fermion parity $(-1)^{\hat F}$
for odd $N_c$
is known from the global-symmetry structure of
QCD~\cite{DumitrescuHsin:2024}.

The novel step is to use this identity to connect the imaginary
angular velocity and the imaginary quark chemical potential exactly at
the density-operator level.
Inserted into \eqref{eq:imaginary-rotation-C:partition-function}, the
identity \eqref{eq:imaginary-rotation-C:operator-identity} yields
\begin{align}
\hat\rho(\tilde\Omega_I+2\pi,\theta_q)
&=\hat\rho(\tilde\Omega_I,\theta_q+\pi),
\label{eq:imaginary-rotation-C:mixed-operator-periodicity}\\
Z(\tilde\Omega_I+2\pi,\theta_q)
&=Z(\tilde\Omega_I,\theta_q+\pi).
\label{eq:imaginary-rotation-C:spectral-flow}
\end{align}
We refer to \eqref{eq:imaginary-rotation-C:mixed-operator-periodicity}
and \eqref{eq:imaginary-rotation-C:spectral-flow} as the mixed
periodicity.

On color-singlet physical states one may write $\hat Q=N_c\hat B$ in
terms of the baryon number operator $\hat B$.
For even $N_c$, $\mathrm e^{\mathrm i\pi\hat Q}=1$ and
\eqref{eq:imaginary-rotation-C:spectral-flow} reduces to a $2\pi$
periodicity in $\tilde\Omega_I$.
For odd $N_c$, $\mathrm e^{\mathrm i\pi\hat Q}=(-1)^{\hat B}$ and the
$2\pi$ periodicity fails in general; applying
\eqref{eq:imaginary-rotation-C:spectral-flow} twice and using the
$2\pi$ periodicity in $\theta_q$, one finds
\begin{align}
Z(\tilde\Omega_I+4\pi,\theta_q)
=Z(\tilde\Omega_I,\theta_q),
\label{eq:imaginary-rotation-C:four-pi-periodicity}
\end{align}
so the imaginary angular velocity is $4\pi$ periodic.
This $4\pi$ periodicity has appeared in
Refs.~\cite{Chen:2023cjt,Chen:2024tkr,Chernodub:2022fractal,Patuleanu:2025dirac},
in each case as a consequence of the spinor sign flip.
Hereafter we set $N_c=3$.

\emph{Exact correspondence to the Roberge-Weiss point.}---Setting
$\tilde\Omega_I=\theta_q=0$ in
\eqref{eq:imaginary-rotation-C:mixed-operator-periodicity} gives
\begin{align}
\hat\rho(2\pi,0)=\hat\rho(0,\pi),
\label{eq:imaginary-rotation-C:RW-operator-mapping}
\end{align}
so the system at $\tilde\Omega_I=2\pi$ shares the expectation values
of all observables with the system at $\theta_q=\pi$.
$SU(3)$ QCD obeys the RW periodicity
$Z(\tilde\Omega_I,\theta_q)=Z(\tilde\Omega_I,\theta_q+2\pi/3)$
~\cite{Roberge:1986mm}, and $\theta_q=\pi$ lies on the RW transition
lines $\theta_q=(2k+1)\pi/3$\;($k\in\mathbb Z$): it is itself an RW
point.

The system at $\theta_q=\pi$ possesses charge conjugation symmetry $C$
exactly: since the eigenvalues of $\hat Q$ are integers,
$\hat C\hat\rho(0,\pi)\hat C^{-1}=\hat\rho(0,\pi)$.
Let $n_I\coloneqq-\mathrm i n_q/T^3$ denote the imaginary quark number
density, with $n_q$ the quark number density.
Because $C$ exchanges a test quark with a test antiquark, the
fundamental Polyakov loop transforms as $C\colon L\mapsto L^{\ast}$ and
$n_I$ as $C\colon n_I\mapsto -n_I$; both $\Im L$ and $n_I$ are thus $C$-odd
order parameters for the breaking of $C$.

Physical-point $N_f=2+1$ lattice QCD extrapolated to the continuum
limit gives the RW endpoint temperature
$T_{\rm RW}=208(5)\,\mathrm{MeV}$ and places the endpoint in the
three-dimensional Ising universality class~\cite{Bonati:2016RW}.
For $T>T_{\rm RW}$, lattice QCD has established that the $\mathbb Z_2$
symmetry is spontaneously broken at the RW point
~\cite{DElia:2009qz,Bonati:2010gi,Bonati:2016RW}.
At $\theta_q=\pi$ this $\mathbb Z_2$ is nothing but $C$; in the broken
phase two equilibrium states coexist, and $\Im L$ and $n_I$ take values
of equal magnitude and opposite sign in the two states.

From now on we consider $T>T_{\rm RW}$.
Identity \eqref{eq:imaginary-rotation-C:operator-identity} rests only
on spin-statistics and $\hat F\equiv\hat Q\pmod 2$; both hold in the
sector with a static fundamental source on the axis---the Wilson line
carries no spin---so the Polyakov loop obeys the same mapping.
Since \eqref{eq:imaginary-rotation-C:RW-operator-mapping} implies
\begin{align}
L\big|_{(2\pi,0)}
&=L\big|_{(0,\pi)},
&n_I\big|_{(2\pi,0)}
&=n_I\big|_{(0,\pi)},
\label{eq:imaginary-rotation-C:RW-observable-mapping}
\end{align}
$C$ is spontaneously broken at $\tilde\Omega_I=2\pi$ as well.
This conclusion depends on no model and no approximation.

This correspondence must not, however, be extended offhand to a finite
interval around $\tilde\Omega_I=2\pi$.
The operator $\hat U$ rotating the system by $\pi$ about an axis
perpendicular to the $z$ axis reverses $\hat J_z$ and commutes with
$\hat H$ and $\hat Q$, so repeated use of
\eqref{eq:imaginary-rotation-C:mixed-operator-periodicity} gives, for
any $n\in\mathbb Z$ and $\delta\in\mathbb R$,
\begin{align}
\hat\rho(2\pi n-\delta,0)
=\hat U
\hat\rho(2\pi n+\delta,0)
\hat U^{-1}.
\label{eq:imaginary-rotation-C:mirror-relation}
\end{align}
Every $\tilde\Omega_I=2\pi n$ is thus the center of a mirror relation
valid throughout space, but corresponds to an RW point only for odd
$n$, and the exact correspondence holds only at $\delta=0$: a deviation
$\delta\neq0$ inserts the statistical weight
$\mathrm e^{\mathrm i\delta\hat J_z}$, and since the angular momentum
of thermal modes grows with the distance from the axis
($J_z\sim rT$), this weight is perturbative only within
$RT\abs{\delta}\lesssim1$, a window that closes in the thermodynamic
limit~\cite{Chernodub:2022fractal,Patuleanu:2025dirac}.
Whether $C$ is broken in a finite range of imaginary angular velocities is
therefore a separate question, which we now address with the one-loop
effective potential.

\emph{One-loop effective potential on the rotation axis.}---To explore
the phase structure at general $\tilde\Omega_I$, we minimize the
one-loop effective potential on the rotation axis $r=0$ globally over
the entire $SU(3)$ Weyl alcove.
Writing the thermal holonomy as
\begin{align}
P=\diag\bigl(
\mathrm e^{\mathrm i\phi_1},
\mathrm e^{\mathrm i\phi_2},
\mathrm e^{\mathrm i\phi_3}
\bigr),
\label{eq:imaginary-rotation-C:polyakov-background}
\end{align}
with $\phi_1+\phi_2+\phi_3\equiv 0\pmod{2\pi}$, we define the
fundamental Polyakov loop by
$L\coloneqq\tfrac{1}{3}\tr P
=\tfrac{1}{3}\sum_{a=1}^{3}\mathrm e^{\mathrm i\phi_a}$.
The free energy per degree of freedom of a massless boson that acquires
the phase factor $\mathrm e^{\mathrm i\varphi}$ upon a full turn of
Euclidean time is~\cite{Gross:1980br,Weiss:1980rj}
\begin{align}
\dfrac{f_B(\varphi)}{T^4}
=\dfrac{\pi^2}{3}
B_4\biggl(
\biggl(\dfrac{\varphi}{2\pi}\biggr)_{\dmod1}
\biggr),
\label{eq:imaginary-rotation-C:boson-free-energy}
\end{align}
where $B_4(x)=x^2(1-x)^2-1/30$.

A gluon in the off-diagonal color component $(a,b)$ with transverse
polarization $s=\pm1$ acquires the phase
$\varphi=\phi_a-\phi_b+s\tilde\Omega_I$, and a diagonal component the
phase $\varphi=s\tilde\Omega_I$.
The gluonic one-loop contribution on the rotation axis is therefore
\begin{align}
\dfrac{V_g}{T^4}
=4\dfrac{f_B(\tilde\Omega_I)}{T^4}
+2\sum_{1\leqslant a<b\leqslant3}
\sum_{s=\pm1}
\dfrac{
f_B\bigl(
\phi_a-\phi_b+s\tilde\Omega_I
\bigr)
}{T^4}.
\label{eq:imaginary-rotation-C:gluon-potential}
\end{align}
The first term is independent of the Polyakov loop background, so the
minima are determined by the second.

A quark flavor of mass $m_f$, for which we set $y_f\coloneqq m_f/T$,
contributes on the rotation axis only through the total angular
momentum modes $j_z=\pm1/2$.
The rotation therefore enters through the statistical weight
$\mathrm e^{\pm\mathrm i\tilde\Omega_I/2}$, and the one-loop
contribution of one flavor reads
\begin{align}
\dfrac{V_f}{T^4}
&=-\dfrac{2y_f^2}{\pi^2}
\sum_{n=1}^{\infty}
\dfrac{(-1)^{n+1}K_2(ny_f)}{n^2}
\cos\biggl(
\dfrac{n\tilde\Omega_I}{2}
\biggr)
\notag\\
&\qquad\times
\sum_{a=1}^{3}
\cos\bigl[
n(\phi_a+\theta_q)
\bigr],
\label{eq:imaginary-rotation-C:quark-potential}
\end{align}
where $K_2$ is the modified Bessel function of the second kind; in
the massless limit $y_f\to0$, $y_f^2K_2(ny_f)\to2/n^2$.
The full one-loop effective potential is then
\begin{align}
V=V_g+\sum_{f=1}^{N_f}V_f.
\label{eq:imaginary-rotation-C:full-potential}
\end{align}
The half-angle factor $\cos(n\tilde\Omega_I/2)$ is the one-loop
fingerprint of the exact mixed periodicity: under
$\tilde\Omega_I\to\tilde\Omega_I+2\pi$ it produces $(-1)^n$, acting
exactly as $\theta_q\to\theta_q+\pi$ does, so
\eqref{eq:imaginary-rotation-C:spectral-flow} is reproduced term by
term in \eqref{eq:imaginary-rotation-C:quark-potential}.

\emph{Two types of phase transition.}---In
the one-loop analysis below, we set $\theta_q=0$, take as the Weyl
alcove the triangle $\phi_1+\phi_2+\phi_3=0$,
$\phi_1\geqslant\phi_2\geqslant\phi_3$,
$\phi_1-\phi_3\leqslant2\pi$, and minimize
\eqref{eq:imaginary-rotation-C:full-potential} numerically over this
entire domain; we also set $t\coloneqq\tilde\Omega_I/\pi$.

\begin{figure}[t]
\centering
\includegraphics[width=\columnwidth]{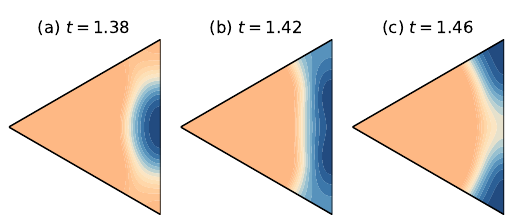}
\caption{
One-loop effective potential on the rotation axis at $\theta_q=0$ for
three massless flavors, over the $SU(3)$ Weyl alcove (orthonormal
Cartan basis; darker shading is lower).
The left vertex is $P=\bm1_3$, the right vertices are the nontrivial
center elements $P=\mathrm e^{\pm 2\pi\mathrm i/3}\bm1_3$, and charge
conjugation flips the triangle upside down.
The global minimum (a) sits at the $C$-symmetric midpoint of the
center-center edge ($t=1.38$), (b) splits into two edge points
exchanged by $C$ ($t=1.42$), and (c) reaches the center-element
vertices ($t=1.46$).}
\label{fig:imaginary-rotation-C:alcove-panels}
\end{figure}

Figure~\ref{fig:imaginary-rotation-C:alcove-panels} displays the
potential over the alcove for three massless flavors.
As $t$ increases, the global minimum moves along the edge of the alcove
that connects the two nontrivial center elements; we therefore study
the potential on this single edge and determine analytically where the
global minimum changes.
At small $\tilde\Omega_I$ a different structure of minima appears,
discussed later.

A point on the edge connecting the two nontrivial center elements can be
parametrized as
\begin{align}
(\phi_1,\phi_2,\phi_3)
=(\pi u,
\pi(2-2u),
\pi(u-2)),
\qquad
\dfrac{2}{3}
\leqslant
u\leqslant
\dfrac{4}{3},
\label{eq:imaginary-rotation-C:C-breaking-line}
\end{align}
where $u=2/3$ and $u=4/3$ correspond to
$\mathrm e^{2\pi\mathrm i/3}\bm1_3$ and
$\mathrm e^{-2\pi\mathrm i/3}\bm1_3$, respectively, and $u=1$ gives
$P=\diag(-1,1,-1)$.
Charge conjugation $C\colon P\mapsto P^{\ast}$ acts on this edge as
$C\colon u\mapsto2-u$, so $u=1$ is the $C$-symmetric point and the
displacement $x\coloneqq1-u$ transforms as $C\colon x\mapsto-x$; we may thus
restrict our attention to $x\geqslant0$.

Substituting \eqref{eq:imaginary-rotation-C:C-breaking-line} into
\eqref{eq:imaginary-rotation-C:full-potential} and subtracting the value
at the $C$-symmetric point $x=0$, we obtain
\begin{align}
\dfrac{V(x)-V(0)}{\pi^2T^4}
=a(t)x^2+9x^4,
\label{eq:imaginary-rotation-C:landau-potential}
\end{align}
for $x\leqslant\min(6-3t,4t-4)/12$, with
$a(t)\coloneqq(27t^2-132t+132)/4$.
Since the quartic term is positive, $x=0$ turns from a minimum into a
maximum once $a(t)$ becomes negative, and two minima exchanged by $C$
appear at $x=\pm x_{\rm min}$ with $x_{\rm min}^2=-a(t)/18$; $a(t)=0$
thus marks the stability boundary of the $C$-symmetric point.
These edge minima are numerically confirmed to be global; hence at the
smaller root of $a(t)$,
\begin{align}
t_C=\dfrac{\tilde\Omega_{I,C}}{\pi}
=\dfrac{22-2\sqrt{22}}{9}
=1.4021\ldots,
\label{eq:imaginary-rotation-C:critical-angle-Nf3}
\end{align}
a second-order transition occurs and $C$ is spontaneously broken.
For $t\gtrsim t_C$ the square-root laws
$x_{\rm min}\propto(t-t_C)^{1/2}$ and
$\Im L(\pm x_{\rm min})\propto\pm(t-t_C)^{1/2}$ follow.

Around the center element at $x=1/3$, the same substitution, written
in terms of the displacement $y\coloneqq1/3-x$, gives
\begin{align}
\dfrac{V(x)-V(1/3)}{\pi^2T^4}
=b(t)y^2-2y^3+9y^4,
\label{eq:imaginary-rotation-C:vertex-landau-potential}
\end{align}
for $y\leqslant\min(3t-2,8-4t)/12$, with
$b(t)\coloneqq(27t^2-36t-4)/4$.
Past the second-order transition, the global minimum becomes degenerate
with the nontrivial center elements
$P=\mathrm e^{\pm2\pi\mathrm i/3}\bm 1_3$ at
$u_{\rm lock}=7/9$, i.e., $x_{\rm lock}=2/9$.
Accordingly, at
\begin{align}
t_{\rm lock}
=\dfrac{\tilde\Omega_{I,{\rm lock}}}{\pi}
=\dfrac{2}{3}
+\dfrac{2\sqrt{111}}{27}
=1.4471\ldots,
\label{eq:imaginary-rotation-C:center-sector-transition}
\end{align}
a first-order transition follows.
Beyond it, $u=2/3$ and $x=1/3$, so just before and after this
transition the order parameter takes the values
$\Im L(2/9)=0.757\ldots$ and $\Im L(1/3)=\sqrt{3}/2=0.866\ldots$.

Numerical minimization over the entire alcove shows that the global
minimum lies on the edge just before lock-in and at the center element
$P=\mathrm e^{2\pi\mathrm i/3}\bm1_3$ just after.
Hence at \eqref{eq:imaginary-rotation-C:center-sector-transition}, where
two distant global minima share the same free energy, a first-order
transition of the on-axis potential occurs and the global minimum jumps
by $x\colon 2/9\to1/3$.
By the mirror relation \eqref{eq:imaginary-rotation-C:mirror-relation}
centered at $\tilde\Omega_I=2\pi$, the same first-order transition also
exists on the opposite side at
$\tilde\Omega_I=4\pi-\tilde\Omega_{I,{\rm lock}}$; between the two
first-order points the two nontrivial center elements
$P=\mathrm e^{\pm 2\pi\mathrm i/3}\bm1_3$ are degenerate global minima
exchanged by charge conjugation.
We refer to this first-order transition, at which the global minimum
moves discontinuously to a center element, as lock-in.

\begin{figure}[t]
\centering
\includegraphics[width=\columnwidth]{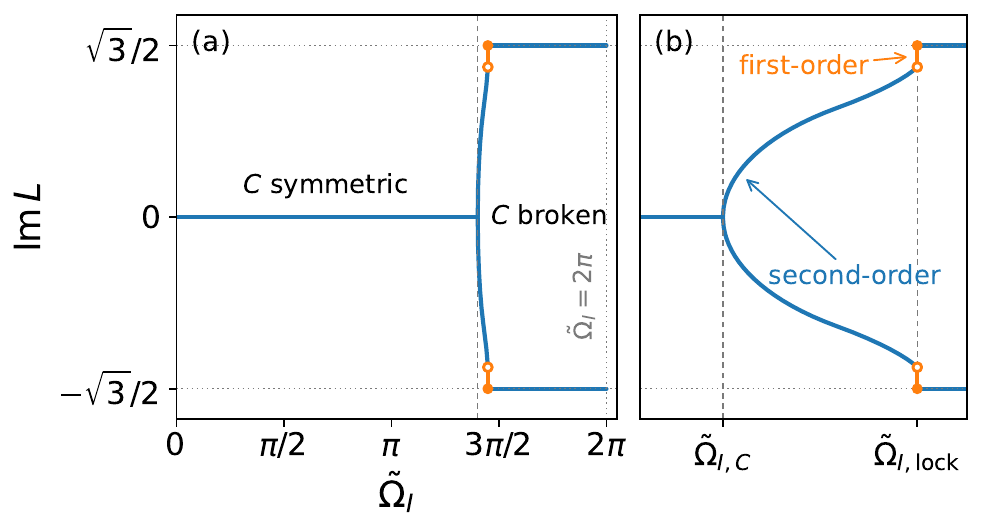}
\caption{
$\Im L$ from the minima of the on-axis one-loop potential in
$2+1$-flavor QCD with $m_s/T=0.30$.
(a) Wide range: the global minimum is uniquely $C$ symmetric below
$\tilde\Omega_{I,C}$, splits continuously into two $C$-conjugate
minima at the second-order transition, and after lock-in
$\Im L=\pm\sqrt{3}/2$ persists up to $\tilde\Omega_I=2\pi$.
(b) Near the two transitions: open and filled circles mark $\Im L$
just before and after the first-order transition;
the dashed verticals mark $\tilde\Omega_{I,C}$ and
$\tilde\Omega_{I,{\rm lock}}$, within $0.1\%$ of the massless results
\eqref{eq:imaginary-rotation-C:critical-angle-Nf3} and
\eqref{eq:imaginary-rotation-C:center-sector-transition}.
}
\label{fig:imaginary-rotation-C:C-order-parameter}
\end{figure}

Finite quark masses are included by evaluating
\eqref{eq:imaginary-rotation-C:quark-potential} numerically.
For massless $u,d$ quarks and an $s$ quark with $m_s/T=0.30$, tracking
the change of the global minimum on the Weyl-alcove edge gives the
second-order and first-order transition points of the one-loop
effective potential on the rotation axis,
$\tilde\Omega_{I,C}/\pi=1.4013\ldots$ and
$\tilde\Omega_{I,{\rm lock}}/\pi=1.4483\ldots$,
each within $0.1\%$ of the massless results.
For the physical strange-quark mass, $T>T_{\rm RW}$ implies
$m_s/T\lesssim0.45$; over $0\leqslant m_s/T\leqslant0.5$ the two
points shift by at most $0.15\%$ and $0.21\%$, so both predictions
are robust throughout the physically relevant temperature range.
The behavior of the $C$-odd order parameter $\Im L$ over
$0\leqslant\tilde\Omega_I\leqslant 2\pi$ is shown in
Fig.~\ref{fig:imaginary-rotation-C:C-order-parameter}.
The one-loop analysis above gives the phase structure of $C$ breaking
on the rotation axis; the exact mapping at $\tilde\Omega_I=2\pi$, by
contrast, holds for the entire system including the off-axis region.

\emph{Accidental degeneracy in the massless case.}---In a certain range of
imaginary angular velocities the massless
one-loop approximation does not decide whether $C$ is realized.
This phenomenon disappears at finite quark masses and is distinct from
the spontaneous $C$ breaking above.
For three massless flavors, the potential
\eqref{eq:imaginary-rotation-C:full-potential} near $P=\bm 1_3$ is
isotropic in the Cartan plane at this order: its minima form a circle,
exactly degenerate for
$2\pi/3<\tilde\Omega_I<(12-2\sqrt{3})\pi/11$, which numerical
minimization over the entire Weyl alcove shows to be global.
The circle contains both $C$-symmetric configurations with $\Im L=0$
and $C$-breaking ones with $\Im L\neq0$, so the massless one-loop
potential alone cannot decide which is selected.
A finite $s$-quark mass does: in $2+1$-flavor QCD with $m_s/T=0.30$,
over the range explored up to $\tilde\Omega_{I,C}$, the global minimum
is unique and satisfies $\Im L=0$
(Fig.~\ref{fig:imaginary-rotation-C:C-order-parameter}).

The same continuous degeneracy occurs in pure $SU(3)$ Yang-Mills
theory, on a circle exactly degenerate for
$(1-1/\sqrt{3})\pi<\tilde\Omega_I<(12-2\sqrt{6})\pi/15$; there no
quark mass is available, and how the degeneracy is resolved cannot be
decided at one loop.
Above this window, however, the one-loop answer is unambiguous: at
$\tilde\Omega_I=\pi/2$ a first-order transition takes the global
minimum to the center-symmetric configuration $L=0$, reproducing the
perturbative confinement under imaginary angular velocity of
Ref.~\cite{Chen:2022smf}.

In full QCD this accidental degeneracy is unrelated to the spontaneous
$C$ breaking at $\tilde\Omega_{I,C}<\tilde\Omega_I$, which is driven
by the quark contribution and survives at finite quark masses; in pure
Yang-Mills theory its resolution remains an open question beyond one loop.

\begin{figure}[t]
\centering
\includegraphics[width=\columnwidth]{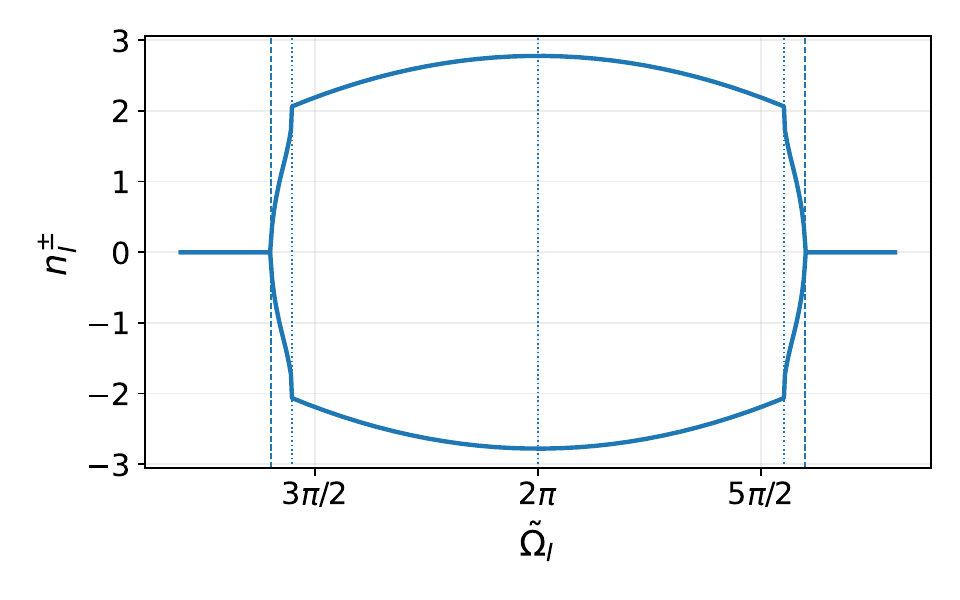}
\caption{
On-axis imaginary quark number density in $2+1$-flavor QCD with
$m_s/T=0.30$, focusing on the $C$-broken window centered at
$\tilde\Omega_I=2\pi$: the two branches $n_I^{\pm}$,
Eq.~\eqref{eq:imaginary-rotation-C:imaginary-density}, evaluated at
the two $C$-conjugate global minima of the broken phase.
The branches carry equal magnitudes and opposite signs; they vanish
in the $C$-symmetric phase, where the global minimum is unique, vary
continuously up to lock-in, and jump there.
}
\label{fig:imaginary-rotation-C:observables}
\end{figure}

\emph{Imaginary quark number density.}---The spontaneous breaking of
$C$ shows up not only in $\Im L$ but in $n_I$.
For the global minima $x=\pm x_{\rm min}$ exchanged by $C$, we define
\begin{align}
n_I^{\pm}
\coloneqq
-\dfrac{\mathrm i n_q^{\pm}}{T^3}
=\dfrac{\partial}{\partial\theta_q}
\dfrac{V(x,\theta_q)}{T^4}
\bigg|_{\theta_q=0,\;x=\pm x_{\rm min}}.
\label{eq:imaginary-rotation-C:imaginary-density}
\end{align}
Figure~\ref{fig:imaginary-rotation-C:observables} shows the two
branches $n_I^{\pm}$ as functions of the imaginary angular velocity
for $2+1$-flavor QCD with $m_s/T=0.30$; for massless quarks, the one-loop
potential gives, at the two degenerate minima at
$\tilde\Omega_I=2\pi$,
\begin{align}
L&=\mathrm e^{\pm 2\pi\mathrm i/3},
&n_I^{\pm}&=\mp\dfrac{8\pi N_f}{27}.
\label{eq:imaginary-rotation-C:RW-order-parameters}
\end{align}
The density $n_I$ is $C$ odd and independent of $\Im L$; that imaginary
rotation induces $n_I^{\pm}\neq 0$ at $\theta_q=0$ means the breaking of $C$
also appears as an on-axis thermodynamic response of a conserved charge.

\emph{Lattice tests.}---Imaginarily rotating lattice simulations
already exist, in pure-glue
~\cite{ChernodubGoyMolochkov:2023,BragutaInhomogeneous:2026} and
dynamical-quark~\cite{BragutaDynamical:2023} settings.
Every exact relation above is directly testable: the mixed periodicity
\eqref{eq:imaginary-rotation-C:spectral-flow}, the $4\pi$ periodicity
\eqref{eq:imaginary-rotation-C:four-pi-periodicity}, the mirror
relation \eqref{eq:imaginary-rotation-C:mirror-relation}, and, most
strikingly, the correspondence
\eqref{eq:imaginary-rotation-C:RW-operator-mapping} together with its
observable mapping
\eqref{eq:imaginary-rotation-C:RW-observable-mapping}; the imaginary
quark number density is precisely the standard order parameter in
lattice studies of the RW endpoint~\cite{DElia:2009qz,Bonati:2016RW}.
Since \eqref{eq:imaginary-rotation-C:RW-operator-mapping} holds at
every temperature, the entire line $\tilde\Omega_I=2\pi$ inherits the
$\theta_q=\pi$ phase structure: $C$ broken for $T>T_{\rm RW}$, restored
below, with a three-dimensional Ising endpoint at physical quark
masses~\cite{DElia:2009qz,Bonati:2010gi,Bonati:2016RW}.
Near $\tilde\Omega_I\simeq1.4\pi$, the one-loop analysis singles out
the $C$-odd $\Im L$: there the breaking is driven by the quark
contribution \eqref{eq:imaginary-rotation-C:quark-potential}, which
the quenched approximation discards---its one-loop potential is that
of pure Yang-Mills theory, with the global minimum at $L=0$
~\cite{Chen:2022smf}---so dynamical and quenched simulations at the same
$\tilde\Omega_I$ should differ sharply in $\Im L$.

\begin{figure}[t]
\centering
\includegraphics[width=\columnwidth]{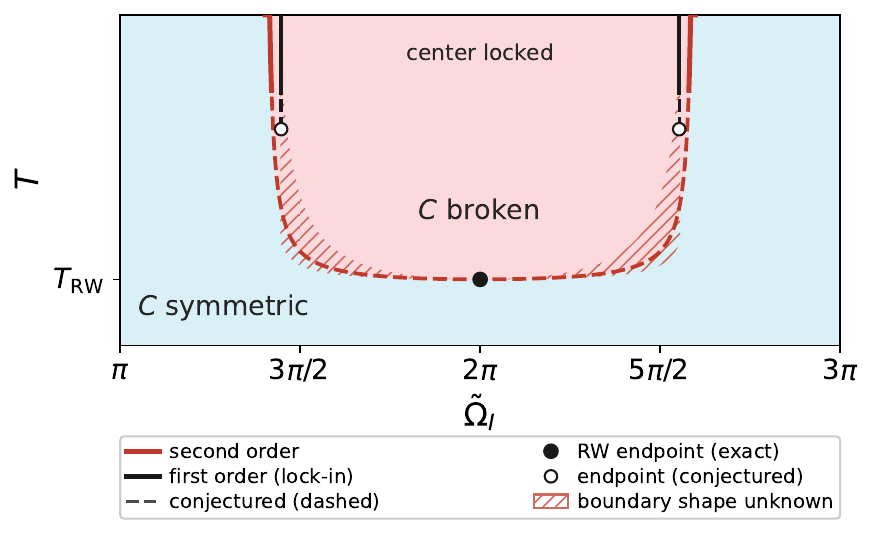}
\caption{Conjectured phase diagram of $C$ breaking on the rotation
axis (schematic).
Solid lines are the one-loop transitions
\eqref{eq:imaginary-rotation-C:critical-angle-Nf3} and
\eqref{eq:imaginary-rotation-C:center-sector-transition} and their
mirrors under \eqref{eq:imaginary-rotation-C:mirror-relation}; the
filled circle is the RW endpoint
$(\tilde\Omega_I,T)=(2\pi,T_{\rm RW})$ inherited exactly through
\eqref{eq:imaginary-rotation-C:RW-operator-mapping}.
The dashed boundary is fixed only where it joins the solid lines and
the endpoint; its course in between is conjectural, and the lock-in
line may terminate at a critical endpoint (open circles).
Mirror symmetry \eqref{eq:imaginary-rotation-C:mirror-relation} makes
the boundary symmetric about $\tilde\Omega_I=2\pi$.
}
\label{fig:imaginary-rotation-C:phase-conjecture}
\end{figure}

Taken together, these tests would chart the conjectured phase diagram
of Fig.~\ref{fig:imaginary-rotation-C:phase-conjecture}: the exactly
mapped line $\tilde\Omega_I=2\pi$ with its endpoint, the mirror
symmetry \eqref{eq:imaginary-rotation-C:mirror-relation}, and the
on-axis transitions
\eqref{eq:imaginary-rotation-C:critical-angle-Nf3} and
\eqref{eq:imaginary-rotation-C:center-sector-transition} fix its
anchors, while the boundary shape and the lock-in line's fate are
nonperturbative and remain for lattice QCD to determine.

\emph{Outlook.}---Higher-order corrections may shift the transition points
\eqref{eq:imaginary-rotation-C:critical-angle-Nf3} and
\eqref{eq:imaginary-rotation-C:center-sector-transition} and lift the
accidental degeneracy of the massless one-loop approximation;
nonperturbative effects become important as $T$ decreases toward
$T_{\rm RW}$. The one-loop analysis is restricted to the rotation axis---away
from it the potential depends on $\tilde
r=rT$~\cite{Chen:2022smf,Chen:2024tkr}---so, while
\eqref{eq:imaginary-rotation-C:RW-operator-mapping} holds system-wide,
the radial distribution of $C$ breaking remains open.
The analytic continuation to real rotation,
$\Omega_I\to -\mathrm i\Omega$, faces the causality
constraint~\cite{Chen:2022smf} and the fractal $\tilde\Omega_I$
dependence in the thermodynamic limit
~\cite{Chernodub:2022fractal,Patuleanu:2025dirac}, so its validity
range is nontrivial.
In summary, one full imaginary turn, $\tilde\Omega_I=2\pi$, carries
hot QCD exactly onto an RW point with spontaneously broken $C$, and
already before the full turn, the one-loop potential predicts, in
closed form, a second-order transition that breaks $C$ and a
first-order transition that locks the minimum into the center
elements.

\bibliography{main}

\end{document}